\documentstyle[epsfig,aps,prl,epsf,multicol]{revtex}

\begin{document}


\hsize\textwidth\columnwidth\hsize
\csname@twocolumnfalse\endcsname

\title{Vortex lattice dynamics in a-NbGe detected by \\
mode-locking experiments}

\author{R.~Besseling, O.~Benningshof, N.~Kokubo and P.H.~Kes}

\address{Kamerlingh Onnes Laboratorium, Leiden
University, P.O. Box 9504, 2300 RA Leiden, the Netherlands.}
\date{\today}
\maketitle

\begin{abstract}
We observed mode-locking (ML) of rf-dc driven vortex arrays in a
superconducting weak pinning a-NbGe film. The ML voltage shows the
expected scaling $V\propto f\sqrt{B}$ with $f$ the rf-frequency
and $B$ the magnetic field. For large dc-velocity (corresponding
to a large ML frequency), the ML current step width exhibits a
squared Bessel function dependence on the rf-amplitude as
predicted for ML of a lattice moving elastically through a random
potential.
\end{abstract}
\begin{multicols}{2}
\narrowtext \noindent

\section{Introduction}

An important tool to study an interacting particle system moving
over a (random) pinning potential, e.g. a sliding solid, charge
density wave or vortex array, is to probe the force-velocity
characteristics with a superimposed rf-drive. For a coherently
(elastically) moving array, the rf-drive may interfere with
lattice modes at an integer $q$ times the 'washboard' frequency
$f_{w}=v/a$, with $a$ the periodicity of the array and $v$ its
velocity, when the frequency $f$ of the rf-drive equals an integer
fraction $1/p$ of $q f_{w}$ \cite{SchmidHaugJLTP73,Fiory}. In the
particular case of a vortex array (VA), the resulting
'mode'-locking (ML) at the velocity $v_{p,q}=(p/q)f a$ appears as
voltage 'plateaus' in the dc current-voltage ($IV$) curves (peaks
in the differential conductance $dI/dV$). For a {\em plastically}
moving VA, which may arise for strong random pinning, at small
velocity or in the vortex liquid phase, the coherent modes are
destroyed and the ML effect vanishes \cite{KoltonshapPRL01}. This
distinction allows to construct dynamic phase diagrams from the
velocity or magnetic field dependence of ML, as we have shown
recently for confined VA's in disordered channels
\cite{Besseling_vcTBchan}.

Focussing on a disordered film with an elastically moving VA (at
relatively large velocity), Ref. \cite{SchmidHaugJLTP73} predicts
a particular form for the dc-current range $\Delta I$ over which
the ML effect occurs. To first approximation, the current range
$\Delta I_{1,1}$ of the main ML step ($p=q=1$) is:
\begin{equation}
\Delta I_{1,1}=2I_c [J_1(I_{rf}R_f/V_{1,1})]^2,
\label{stepwidthformula}
\end{equation}
where $J_1(x)$ is the first order Bessel function, $R_f$ the flux
flow resistance, $V_{1,1}$ the fundamental ML voltage and $I_c$
the dc-critical current in absence of rf-drive.

So far, the only report of ML of a VA in a disordered
superconducting film is Ref. \cite{Fiory}, where Al films were
used and the rf-amplitude dependence was not addressed. In this
report we show that a-NbGe films with weak random pinnning
\cite{Kes,Berghuis} also exhibit the ML effect and we examine its
dependence on $I_{rf}$.

\section{Experimental}

A $400$ nm thick amorphous (a-)NbGe strip ($T_c\simeq 3.0$ K and
normal state resistivity $\rho_n\simeq 2$ $\mu\Omega$m), with
current leads covered by a strong pinning NbN film ($T_c\simeq 10$
K) to reduce sample impedance, was prepared by RF sputtering and
standard lithography. In the resulting four-probe configuration
both the strip width and the distance between voltage contacts was
$100$ $\mu$m. In the transport measurement, the rf-current
(amplitude ranging from $0-0.5$ mA) was applied through a matched
circuit. All data were taken with the sample immersed in
superfluid $^4$He at a temperature $T=1.8$ K and the magnetic
field $B$ perpendicular to the film.

\section{Results and discussion}

\begin{figure}
\epsfig{file=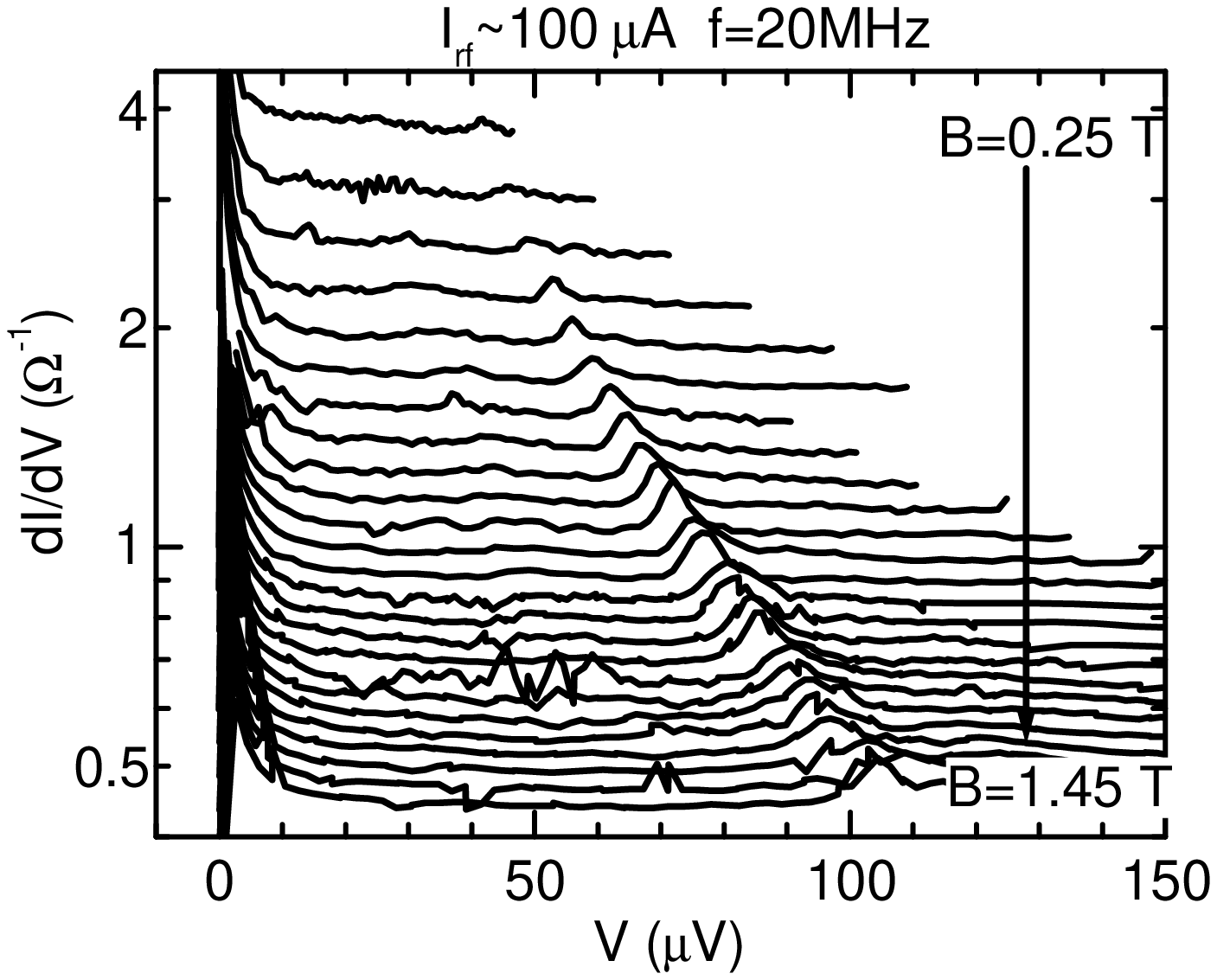, width=5cm, height=4cm} \vspace{0.5cm}
\caption{Differential conductance $dI/dV$ versus dc-voltage $V$ in
presence of a $20$ MHz rf-current of amplitude $I_{rf} \simeq 100$
$\mu$A for fields ranging from $B=0.25$ T to $1.45$ T with steps
of $50$ mT.} \label{fig1}
\end{figure}

In Fig. \ref{fig1} we plot $dI/dV$ versus the dc-voltage measured
with a superimposed rf-current of $20$ MHz and an amplitude
$I_{rf}\simeq 100$ $\mu$A at magnetic fields $B$ ranging from
$0.25$ to $1.45$ T. In the whole field range, the curves show a
clear ML peak in $dI/dV$ which can be identified as the
fundamental $p=q=1$. Additionally, for some fields, (weaker)
subharmonic peaks can be observed. The ML voltage $V_{1,1}$ of the
main peak scales with magnetic field as $V \sim \sqrt{B}$, in
agreement with the relation $V\propto v\cdot B=f a B$ where $a
\sim1 /\sqrt{B}$. In addition we checked that the ML voltage at
fixed field scales linearly with the frequency of the applied
rf-current. The observed ML implies that the moving VA
predominantly experiences elastic deformations, validating the
previous use of the collective pinning theory in this material
\cite{Kes}.

\begin{figure}
\epsfig{file=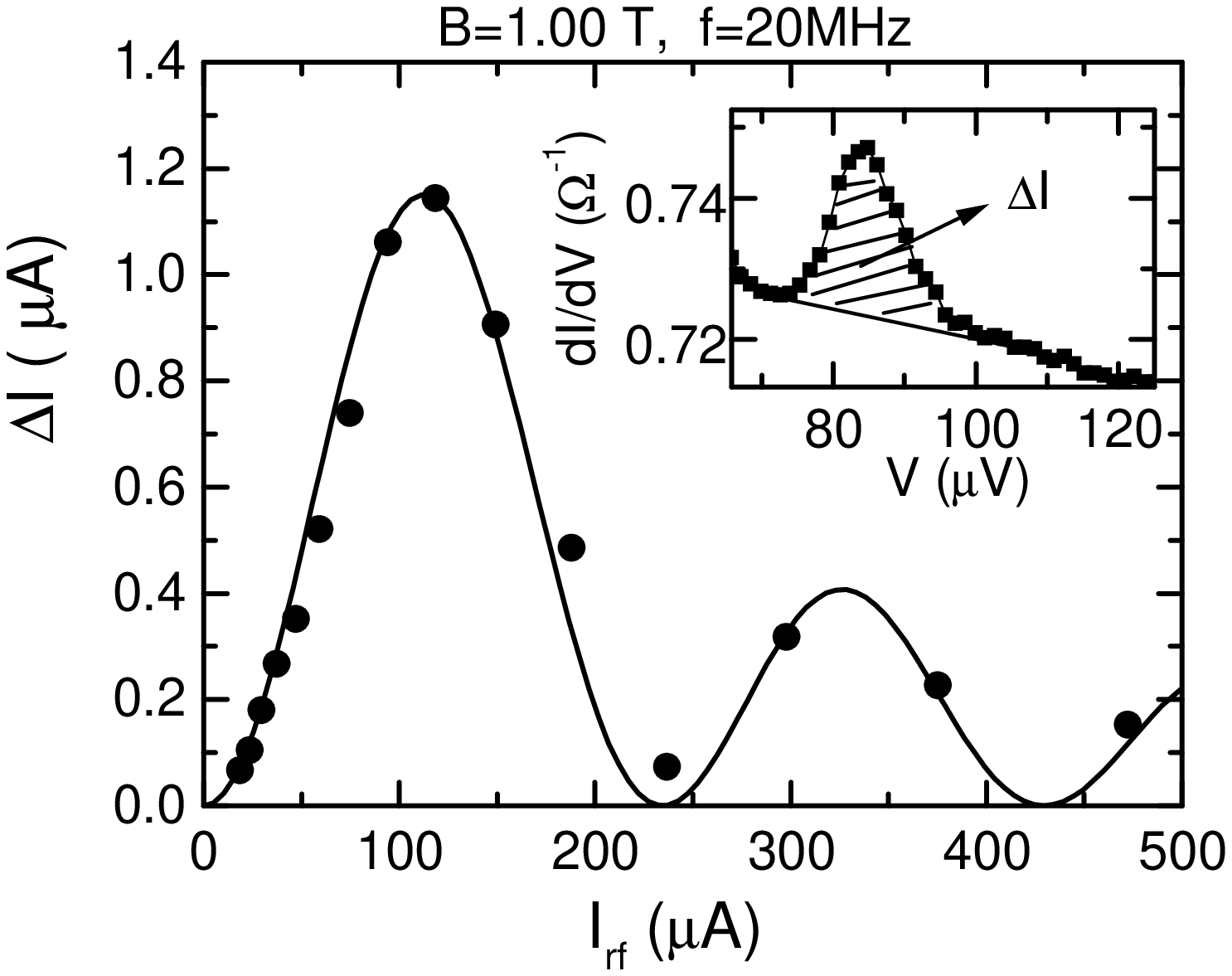, width=5cm, height=4cm} \vspace{0.5cm}
\caption{Data points: ML current width of the fundamental step
$\Delta I_{1,1}$ (as defined in the inset) versus $I_{rf}$ at
$B=1.00$ T and $f=20$ MHz. The drawn line represents Eq.
(\ref{stepwidthformula}) with $I_c=1.7$ $\mu$A as the only fit
parameter.} \label{fig2}
\end{figure}

We now turn to the rf-amplitude dependence of the ML. Due to
rounding of the steps in the $IV$ curves the ML current width
$\Delta I$ is best defined as shown in the inset to Fig.
\ref{fig2}. The data in the main panel show the result of such
analysis at $B=1$ T. The width $\Delta I$ clearly exhibits
oscillatory behavior as function of $I_{rf}$. We also plot the
dependence according to Eq. (\ref{stepwidthformula}) (drawn line)
where both $R_f$ and $V_{1,1}$ were extracted from the experiment,
the only fit parameter being $I_c$. The experimental data follow
the theoretical prediction remarkably well. The fitted value
$I_c=1.7$ $\mu$A is smaller than the critical current $I_{c,L}
\simeq 6$ $\mu$A determined by linearly extrapolating the 'bare'
$IV$ curve (at $I_{rf}=0$) to $V=0$. This difference may arise
from the approximation made in Eq. (\ref{stepwidthformula}) (see
\cite{SchmidHaugJLTP73}). Comparing our result to the numerical
work of Kolton et al. \cite{KoltonshapPRL01}, we find no evidence
for their suggestion that $\Delta I\sim J_1(I_{rf}R_f/V_{1,1})$,
valid for an rf-dc driven Josephson junction at large voltage ( a
single particle moving rapidly in a periodic potential). This
discrepancy with their results, including the lack of step
broadening, might originate from the neglect of long wavelength
fluctuations in a finite size simulation.

Finally we mention that preliminary data taken at a larger
frequency $f=100$ MHz exhibit ML up to a field $B^*(<B_{c2})$ at
which the critical current has completely vanished and the IV
curves have become fully linear with a slope $R_f\simeq 0.75 R_n$.
The field $B^*$ may be identified as the equilibrium melting field
of the 2D vortex lattice, in line with Ref. \cite{Berghuis}. For
fields slightly below $B^*$, ML sensitively depends on the
frequency. Since the VA softens considerably just below melting,
the effect of pinning disorder is stronger in this regime. Based
on the dynamic ordering theory in \cite{Koshelevrecryst} and our
measurements in disordered mesoscopic channels
\cite{Besseling_vcTBchan}, one then indeed expects the appearance
of a coherent state to depend strongly on the vortex velocity.

This work was supported by the 'Stiching voor Fundamenteel
Onderzoek der Materie' (FOM).

\end{multicols}


\begin{references}

\bibitem{SchmidHaugJLTP73}A. Schmid and W. Hauger, J. Low. Temp. Phys.
{\bf 11}, 667 (1973).

\bibitem{Fiory}A. T. Fiory, Phys.\ Rev.\ Lett.\ {\bf 27}, 501 (1971);
Phys.\ Rev.\ B {\bf 7}, 1881 (1973).

\bibitem{KoltonshapPRL01}A.B. Kolton {\it et al.}, Phys.\ Rev.\ lett.\ {\bf 86}, 4112 (2001).

\bibitem{Besseling_vcTBchan}R. Besseling {\it et al.}, cond-mat/$0302187$; N. Kokubo {\it et al.},
Phys.\ Rev.\ Lett.\ {\bf 88}, 247004 (2002).

\bibitem{Kes}P.H. Kes and C.C. Tsuei, Phys. Rev. Lett. {\bf
47}, 1930 (1981).

\bibitem{Berghuis}P. Berghuis {\it et al.}, Phys. Rev. Lett. {\bf 65}, 2583
(1990); Phys.\ Rev.\ B {\bf 47}, 262 (1993).

\bibitem{Koshelevrecryst}A.E. Koshelev and V.M. Vinokur, Phys.\ Rev.\ Lett.\ {\bf 73}, 3580
(1994).

\end{references}
\end{document}